\documentclass{article}[11pt]
\usepackage{amsfonts}
\usepackage{CJK,amsmath,indentfirst}
\usepackage[numbers,sort&compress]{natbib}
\usepackage{graphics}
\usepackage{graphicx}
\begin{document}

\title{\bf Consistent Riccati Expansion and Solvability}

\author{\footnotesize S. Y. Lou$^{1,2}$\\
\footnotesize $^{1}$\it Shanghai Key Laboratory of Trustworthy Computing, East China Normal University, Shanghai 200062, China\\
\footnotesize $^{2}$\it Department of Physics, Ningbo University, Ningbo 315211, China}
\date{}
\maketitle
\parindent=0pt
\textbf{Abstract:} A consistent Riccati expansion (CRE) is proposed for solving nonlinear systems with the help of a Riccati equation. A system is defined to be CRE solvable if it has a CRE. Various integrable systems are CRE solvable. Furthermore, it is also revealed that many CRE solvable systems, including the Korteweg de-Vries, Kadomtsev-Patviashvili, nonlinear Schr\"dinger and sine-Gordon equations, possess a common determining equation which describes interactions between a soliton and a cnoidal wave.

\vskip.4in
\renewcommand{\thesection}{\arabic{section}}
\parindent=20pt

\section{Introduction}

A trouble and tedious but very important problem is to find exact solutions of nonlinear systems. To solve this problem, some elegant approaches such as the inverse scattering transformation \cite{IST}, Darboux-B\"acklund transformations \cite{DT}, and nonlinearizations of Lax pairs\cite{NL} have been established. However, these effective methods are inscrutable or mystical for people who are neither  mathematicians nor theoretical physicists. Thus it is very necessary to establish some types of quite simple and understandable methods to construct exact solutions. For instance, the hyperbolic tangent (tanh) function expansion method \cite{tanh}, the auxiliary equation (say, Riccati equation) method \cite{Ric} and the homogeneous balance method \cite{homo} are usually used to find (quasi-) traveling solitary wave solutions. Nevertheless, it is unfortunate that these feasible methods might lose some essential information of the original nonlinear systems, and consequently, only some quite special solutions can be obtained.
In fact, the simple methods mentioned above can be further developed to find much more general solutions retrieving the missing essential properties. In addition, one can even clarify the integrability of some types of nonlinear systems. In Refs. \cite{Lou1,Lou2,Lou3}, we have generalized the tanh function expansion method to find not only various interaction solutions between different types of excitations but also possible new integrable systems.

In the next section, we propose a consistent Riccati expansion (CRE) method, which can be considered as an extension version of the usual Riccati equation method and the tanh function expansion method, such that we could find new solutions of nonlinear systems and strong signals to clarify possible integrable models. Based on the new method, we define a model is CRE solvable if it has a CRE. The method is systematically illustrated by some important nonlinear systems. Section 3 is devoted to clarifying the possible CRE solvable cases of the fifth order KdV system. In Section 4, it is exhibited that many CRE solvable systems possess a same determining equation which describes interesting interactions between a soliton and a cnoidal wave. The last section is a short summary and discussion.

\section{Consistent Riccati expansion for some nonlinear systems}
For a given derivative nonlinear polynomial system,
\begin{equation}\label{PNLE}
P( {\mbox{\bf x}},\  t,\ u)=0, \quad {\mbox{\bf x}}=\{x_1,\ x_2,\ \ldots,\ x_n\},
\end{equation}
we  aim to look for the following possible truncated expansion solution
\begin{equation}\label{Ep}
u=\sum_{i=0}^n u_i R^i(w),
\end{equation}
where $R(w)$ is a solution of the Riccati equation
\begin{equation}\label{RCT}
R_w=a_0+a_1R+a_2R^2,
\end{equation}
$n$ should be determined from the leading order analysis of \eqref{PNLE}, and all the expansion coefficient functions $u_i$  should be determined by vanishing the coefficients of the like powers of $R(w)$ after substituting \eqref{Ep} with \eqref{RCT} into \eqref{PNLE}. \\
\bf Definition. \rm If the system for $u_i\ (i=0,\ \ldots,\ n)$ and $w$ obtained by vanishing \em all \rm the coefficients of the powers of $R(w)$ after substituting \eqref{Ep} with \eqref{RCT} into \eqref{PNLE} is consistent, or not over-determined, we call the expansion \eqref{Ep} is a \em consistent Riccati expansion \rm (CRE) and the nonlinear system \eqref{PNLE} is defined \em CRE solvable.\rm

In the following, we apply the proposed new method to nine concrete examples, as a result, we can find that a variety of integrable systems are actually CRE solvable.

\bf Example 1. \em The Korteweg de-Vries (KdV) equation. \rm For the KdV equation \cite{KdV}
\begin{equation}\label{KdV}
u_t+6uu_x+u_{xxx}=0,
\end{equation}
the leading order analysis leads to $n=2$. Substituting \eqref{Ep} with \eqref{RCT} and $n=2$ into \eqref{KdV} yields
\begin{eqnarray}\label{KdVa}
0&=& 12a_2w_xu_2(u_2+2a_2^2w_x^2)R^5\nonumber\\
&&+6\left[u_2w_x(2u_2+9a_2^2w_x^2)a_1+a_2^3u_1w_x^3
+3w_xa_2^2(w_xu_2)_x
+3a_2u_1u_2w_x+u_2u_{2x}\right]R^4\nonumber\\
&& +\left[4u_2w_x(3u_2+10a_2^2w_x^2)a_0
+38a_2a_1^2u_2w_x^3+6a_2^2w_x(u_1w_x)_x
+6a_1w_x\big[3u_1u_2+2a_2^2u_1w_x^2\right.\nonumber\\
&& \left.+5a_2(w_xu_2)_x\big]
+6(u_1u_2)_x+2a_2\big[u_2(w_t+w_{xxx})
+3w_(2u_0u_2+u_1^2)+3(u_{2x}w_x)_x)\big]\right]R^3\nonumber\\
&& +\left[2a_0w_x[2w_x^2a_2(2u_1a_2+13 u_2 a_1)+12 a_2(w_x u_2)_x+9u_1u_2]+8u_2w_x^3a_1^3+a_1^2 w_x[7 u_1 w_x^2 a_2\right.\nonumber\\
&&+12 (w_xu_2)_x]+a_1[9a_2w_x (u_1w_x)_x +12 u_0u_2w_x +6(u_{2x}w_x)_x+2 u_2(w_{xxx}+w_t)+6 u_1^2 w_x]\nonumber\\
&&\left.+a_2[3(u_{1x}w_x)_x+u_1(w_{xxx}+w_t+6u_0w_x]
+u_{2t}+u_{2xxx}+6u_1u_{1x}
+6(u_2u_{0})_x\right]R^2\nonumber\\
&& +\left\{16 a_2 a_0^2 u_2 w_x^3 + a_0\big[14 a_1^2 u_2 w_x^3 +2 a_1 w_x[4 a_2 u_1 w_x^2 +9 (w_x u_2)_x]+6 a_2 w_x (u_1 w_x)_x\right.\nonumber\\
&&+12 u_0 u_2 w_x+6 (u_{2x}w_x)_x+2 u_2 (w_{xxx}+ w_t)+6 u_1^2 w_x\big]+a_1^3 u_1 w_x^3 +3 a_1^2 w_x (u_1 w_x)_x \nonumber\\
&&+a_1[3 (u_{1x} w_x)_x+ u_1 (w_{xxx}+w_t+6 u_0 w_x)] \left.+u_{1t}+u_{1xxx}+6 (u_0 u_1)_x\right\}R\nonumber\\
&&+a_0^2[6 a_1 u_2 w_x^3+2 a_2 u_1 w_x^3 +6 w_x (w_x u_2)_x]+ a_0[a_1^2 u_1 w_x^3 +3 a_1 w_x (u_1 w_x)_x+3 (u_{1x} w_x)_x\nonumber\\
&&+u_1( w_{xxx}+w_t+6 u_0 w_x)]+u_{0t}+6 u_0 u_{0x}+u_{0xxx}.
\end{eqnarray}

Setting zero the coefficients of all the same powers of $R$ in \eqref{KdVa}, we have six over-determined equations for only four undetermined functions $u_0,\ u_1,\ u_2$ and $w$. It is fortunate that for many integrable models, these kinds of over-determined systems may be consistent. In this case, from the coefficients of $R^5,\ R^4$ and $R^3$, we can simply find
\begin{subequations}\label{u012}
\begin{equation}\label{u2}
u_2=-2a_2^2w_x^2,
\end{equation}
\begin{equation}\label{u1}
u_1=-2a_2(a_1w_x^2+w_{xx}),
\end{equation}
\begin{equation}\label{u0}
u_0=-\frac16\frac{w_t}{w_x}
-\frac16(a_1^2+8a_0a_2)w_x^2-a_1w_{xx}
-\frac23\frac{w_{xxx}}{w_x}
+\frac12\frac{w_{xx}^2}{w_x^2}.
\end{equation}
\end{subequations}
The coefficient of $R^2$ in \eqref{KdVa} becomes identically zero by using \eqref{u012}. Then from the coefficient of $R$ in \eqref{KdVa}, we find the $w$ equation
\begin{equation}\label{wt}
w_t=-w_{xxx}+\frac32\frac{w_{xx}^2}{w_x}
+\frac12 \delta w_x^3+\lambda w_x,\ \delta\equiv (a_1^2-4a_0a_2).
\end{equation}
Finally, one can verify that the coefficient of $R^0$ in \eqref{KdVa} is identically zero. Evidently, the condition in the {\bf Definition} is satisfied, and thus the KdV equation \eqref{KdV} is CRE solvable. It is noted that the $w$ equation \eqref{wt} is a generalization of the Schwarzian KdV equation (without $\delta$ term) and the potential modified KdV equation (cancelling the rational term).

In summary, we have the following theorem:

\bf Theorem 1. \rm If $w$ is a solution of Eq. \eqref{wt}, then
\begin{equation}\label{ru_kdv}
u=-\frac{\lambda}6-a_1w_{xx}
-\frac14(a_1^2+4a_0a_2)w_x^2
-\frac12\frac{w_{xxx}}{w_x}+\frac14\frac{w_{xx}^2}{w_x^2}
-2a_2(a_1w_x^2+w_{xx})R-2a_2^2w_x^2R^2
\end{equation}
is a solution of the KdV equation \eqref{KdV} with $R\equiv R(w)$ being a solution of the Riccati equation \eqref{RCT}.

\bf Example 2. \em The Kadomtsev-Petviashvili (KP) equation. \rm Applying the same procedure to the KP equation \cite{KP}
\begin{equation}\label{KP}
(u_t+6uu_x+u_{xxx})_x+\gamma u_{yy}=0,
\end{equation}
we can write down the following theorem. It is noted that the detailed derivation or proof is omitted since the procedure is exactly same to the KdV case, which is also valid for the following theorems for other different models.

\bf Theorem 2. \rm If $w$ is a solution of
\begin{equation}\label{wxt}
\left(C+S
-\frac{\delta} 2 w_x^2+\frac{\gamma}2 K^2\right)_x
+\gamma K_y=0,
\end{equation}
with
\begin{equation}\label{CSK}
C\equiv \frac{w_t}{w_x},\qquad S\equiv\frac{w_{xxx}}{w_x}
-\frac32\frac{w_{xx}^2}{w_x^2}, \qquad K\equiv \frac{w_y}{w_x},
\end{equation}
then
\begin{equation}\label{ru_kp}
u=-\frac16\frac{w_t}{w_x}-a_1w_{xx}
-\frac16(a_1^2+8a_0a_2)w_x^2
-\frac23\frac{w_{xxx}}{w_x}+\frac12\frac{w_{xx}^2}{w_x^2}
-\frac{\gamma}6\frac{w_y^2}{w_x^2}
-2a_2(a_1w_x^2+w_{xx})R-2a_2^2w_x^2R^2
\end{equation}
is a CRE of the KP equation.

\bf Example 3. \em The Boussinesq equation. \rm The theorem for the Boussinesq equation \cite{Bq}
\begin{equation}\label{Bq}
u_{tt}+(6uu_x+u_{xxx})_x=0,
\end{equation}
is stated as follows:

\bf Theorem 3. \rm If $w$ is a solution of
\begin{equation}\label{wtt}
\left(S
-\frac{\delta} 2 w_x^2+\frac12C^2\right)_x
+C_t=0,
\end{equation}
then
\begin{equation}\label{ru_bq}
u=-a_1w_{xx}
-\frac16(a_1^2+8a_0a_2)w_x^2
-\frac23\frac{w_{xxx}}{w_x}+\frac12\frac{w_{xx}^2}{w_x^2}
-\frac16\frac{w_t^2}{w_x^2}
-2a_2(a_1w_x^2+w_{xx})R-2a_2^2w_x^2R^2
\end{equation}
is a CRE of the Boussinesq equation.

\bf Example 4. \em The AKNS (Ablowitz-Kaup-Newell-Segur) system. \rm For the AKNS system \cite{AKNS}
\begin{subequations}\label{AKNS}
\begin{eqnarray}
&& p_t+\frac12ibp_{xx}-ip^2q=0,\qquad b^2=1,\\
&&q_t-\frac12ibq_{xx}+iq^2p=0,
\end{eqnarray}
\end{subequations}
the theorem reads

\bf Theorem 4. \rm If $w$ is a solution of
\begin{equation}\label{w}
C_t+\frac18\left(2S-12C^2-16b\lambda C+\delta w_x^2\right)_x=0,
\end{equation}
then
\begin{subequations}\label{pqw}
\begin{equation}
p=\sqrt{b} \left[a_2w_x R+ib\frac{w_t}{w_x}+i\lambda +\frac12\frac{w_{xx}}{w_x}+\frac12w_x\right]e^{iu},
\end{equation}
\begin{equation}
q=\sqrt{b} \left[a_2w_x R-ib\frac{w_t}{w_x}-i\lambda +\frac12\frac{w_{xx}}{w_x}+\frac12a_1w_x\right]e^{-iu},
\end{equation}
\end{subequations}
is a solution of the AKNS system \eqref{AKNS} with the consistent conditions for the `phase' $u$
\begin{subequations}\label{uw}
\begin{equation}\label{ux}
u_x=2bC+\lambda,
\end{equation}
\begin{equation}\label{ut}
u_t=3bC^2+4\lambda C-\frac14\delta bw_x^2+\frac32 b\lambda^2-\frac b2S.
\end{equation}
\end{subequations}
It is remarkable that the consistent condition $u_{xt}=u_{tx}$ of \eqref{uw} is nothing but \eqref{w}.

\bf Example 5. \em Sine-Gordon (sG) equation. \rm For the sine-Gordon equation \cite{sG}
\begin{equation}\label{sG}
v_{xt}-2m\sin(v)=0,
\end{equation}
in order to use the CRE method, we have to transform it to a derivative polynomial form
\begin{equation}\label{sgu}
u_tu_x-uu_{xt}+mu^3-mu=0,
\end{equation}
through
\begin{equation}\label{uv}
u=\exp(iv).
\end{equation}
For the variant form \eqref{sgu} of the sG equation, we can establish the following theorem:

\bf Theorem 5. \rm If $w$ is a solution of the consistent system
\begin{subequations}\label{sgw}
\begin{eqnarray}
&&(CC_{xt}-C_xC_t)\lambda+mC(C^2-\lambda^2)=0, \\
&&\lambda(\delta C^2w_x^2+2CC_{xx}+2C^2S-C_x^2)-2C^2m=0,
\end{eqnarray}
\end{subequations}
then Eq. \eqref{sgu} has a CRE solution
\begin{eqnarray}\label{ru}
u=\frac2mw_tw_xa_2^2R^2+\frac{2a_2}m(w_{xt}+a_1w_xw_t)R
+\frac{(w_{xt}+a_1w_xw_t)^2}{2mw_xw_t}.
\end{eqnarray}

It should be emphasized that for the sG equation, one $w$ function would satisfy two equations \eqref{sgw}, however we can still call it CRE solvable because those two equations are consistent, i.e., $C_{xtx}=C_{xxt}$ is identically satisfied.

\bf Example 6. \em Modified asymmetric Veselov-Novikov (VN) equation. \rm For the modified asymmetric VN equation \cite{VN}
\begin{subequations}\label{VN}
\begin{eqnarray}
&&u_t-u_{xxx}-3u_xv_x-\frac32uv_{xx}=0,\nonumber\\
&&v_y=u^2,
\end{eqnarray}
\end{subequations}
it is straightforward to find the transformation theorem as follows:

\bf Theorem 6. \rm If $w$ is a solution of
\begin{eqnarray}\label{VNw}
C_y=\frac{\delta}4w_x(w_xK_x+4Kw_{xx})+\frac{3K_x^3}{4K^2}
-\frac3{2K}K_xK_{xx}+\frac12SK_x+KS_x+K_{xxx},
\end{eqnarray}
then the VN system \eqref{VN} possesses the CRE form
\begin{eqnarray}\label{VNuv}
&&u=-\frac{b}2\sqrt{-w_xw_y}\left(a_1+\frac{w_{xy}}{w_xw_y}
+2a_2R\right),\nonumber\\
&&v=v_0-a_2w_xR,
\end{eqnarray}
where $v_0$ is related to $w$ by a consistent system
\begin{eqnarray}\label{v0xy}
&&v_{0x}=\frac{\delta}{12}w_x^2-\frac12a_1w_{xx}
+\frac{C-S}3-\frac{w_{xx}^2}{w_x^2},\nonumber\\
&&v_{0y}=-\frac{K_x^2}{4K}-\frac{K_x}{2w_x}(a_1w_x^2+w_{xx})
+\frac{\delta}4Kw_x^2-\frac{a_1}2Ka_1w_{xx}
-\frac{K}4\frac{w_{xx}^2}{w_x^2}.
\end{eqnarray}
It is noted that the compatibility condition $v_{0xy}=v_{0yx}$ is nothing but the $w$ equation \eqref{VNw}.

\bf Example 7. \em Dispersive water wave (DWW) equation. \rm For the DWW equation \cite{DWW}
\begin{subequations}\label{DWW}
\begin{eqnarray}
&&u_t-(u_{xx}-3vu_x+3uv^2-3u^2)_x=0,\\
&&v_t-(v_{xx}+3vv_x+v^3-6uv)_x=0,
\end{eqnarray}
\end{subequations}
we have the following CRE theorem:

\bf Theorem 7. \rm The DWW equation \eqref{DWW} has a CRE
\begin{eqnarray}\label{WWexp}
&&u=a_2^2w_x^2R^2+a_2(w_{xx}+a_1w_x^2)R
+\frac12(a_1+b\sqrt{-\delta})w_x+v_{0x}+a_0a_2w_x^2,\nonumber\\
&&v=a_2w_xR+\frac12(a_1+b\sqrt{-\delta})w_x+v_0,\quad b^2=1,
\end{eqnarray}
with $\{w,\ v_0\}$ being a solution of the coupled system
\begin{subequations}\label{wv0}
\begin{eqnarray}
&&w_t=\left(w_{xx}-3v_0w_x\right)_x-w_x\big[\delta w_x^2+3b\sqrt{-\delta}(w_{xx}-v_0w_x)-3v_0^2\big],\label{w0}\\
&&
v_{0t}=(v_{0xx}-3v_0v_{0x}+v_0^3)_x.\label{v0}
\end{eqnarray}
\end{subequations}

It is clear that to prove the CRE solvability we can simply take $v_0=0$. However, nonzero $v_0$ will lead to more exact solutions of DWW system.

\bf Example 8. \em Burgers equation. \rm Here, we write down the CRE theorem for a simple C-integrable model, the Burgers equation
\begin{eqnarray}\label{Be}
u_t=u_{xx}+2uu_x.
\end{eqnarray}

\bf Theorem 8. \rm If $w$ is a solution of
\begin{eqnarray}\label{Bew}
C_t=\left(\frac12C^2-S+2C_x-\frac{\delta}2w_x^2\right)_x,
\end{eqnarray}
then
\begin{eqnarray}\label{Besol}
u=\frac{w_t-w_{xx}}{2w_x}-\frac{a_1}2w_x-a_2w_xR
\end{eqnarray}
is a CRE solution of the Burgers equation \eqref{Be}.

The above examples reveal that for many kinds of integrable models, the truncated expansions based on the Riccati equation will lead to some $w$ equations, which are the generalization of the Schwarzian equations of the original nonlinear systems because they will reduce back to their Schwarzian forms when $\delta=0$.

It is naturally expected that this elegant property will be lost for nonintegrable systems. Here, we just present one non-CRE solvable example.

\bf Example 9. \em Non-integrable KdV-Burgers (KdV-B) equation. \rm For the KdV-B equation
\begin{equation}\label{KdVB}
u_t+6uu_x+\nu u_{xx}+u_{xxx}=0,
\end{equation}
with the same procedure as previous cases, the substitution of the Riccati expansion
\begin{equation}\label{KdVBe}
u=u_0+u_1R+u_2R^2
\end{equation}
into the model will result in
\begin{eqnarray}\label{KdVBu}
u&=&\frac{\nu^2}{150}-\frac15(a_1w+\ln w_x)_x-\frac16 \frac{w_t}{w_x}-\frac16(a_1^2-8a_0a_2)w_x^2-a_1w_{xx}
-\frac23\frac{w_{xxx}}{w_x}+\frac12\frac{w_{xx}^2}{w_x^2}
\nonumber\\
&&
-\frac{2a_2}5(5w_{xx}+\nu w_x+5a_1w_x^2)R-2a_2^2w_x^2R^2.
\end{eqnarray}
However, owing to the nonintegrability of the model, two non-completely consistent $w$ equations have to be introduced as
\begin{subequations}\label{KdVBw}
\begin{equation}\label{Cx}
C_x=-S_x-\delta w_xw_{xx}-\frac{\nu^3}{125}-\frac{\nu}5(2S+\delta w_x^2),
\end{equation}
\begin{eqnarray}\label{Ct}
C_t&=&S_{xxx}+\frac{15}2\delta\frac{w_{xx}^2}{w_x}
-\frac52\nu\delta w_{xx}^2-\frac{\nu}{3125}(50S+\nu^2)(50S-\nu^2+25C)
-\nu S_{xx}\nonumber\\
&&+\frac{\delta}{25}(225S-25C-11\nu^2)w_xw_{xx}
-\frac15\nu\delta(C+9S)w_x^2+\delta^2w_x^3w_{xx}\nonumber\\
&&
-\frac15\nu \delta^2w_x^4 +
\left(2S+2\delta w_x^2-C-\frac{11}{25}\nu^2\right)S_x.
\end{eqnarray}
\end{subequations}
It is noted that  if $\nu=0$, then two equations in \eqref{KdVBw} will be degenerated to one equation \eqref{wt}. While for $\nu\neq0$, the consistent condition $C_{tx}-C_{xt}$ is not identically zero except for
\begin{eqnarray}\label{Sxxx}
&&S_{xxx}+\left(2\delta w_x^2+2S+\frac{\nu^2}{225}\right)S_x +\frac{\nu}{45}\delta^2w_x^4+\delta^2w_x^3w_{xx}
+\frac{19}{45}\delta\nu Sw_{x}^2+\frac{\delta}{225}(\nu^2+2025S)w_xw_{xx}
\nonumber\\
&&\quad +\frac13\nu S_{xx}+\frac56\nu\delta w_{xx}^2+\frac{\nu(50^2S^2-\nu^4)}{28125}
+\frac{15}2\frac{\delta w_{xx}^3}{w_x}=0.
\end{eqnarray}
Consequently, the non-integrable KdV-Burgers equation is non-CRE solvable.

\section{Searching for integrable systems via CRE}

The results concerning the examples in the last section provide us an approach to find CRE solvable systems which may be strongly integrable systems. Now, as an illustration, we try to clarify some possible CRE solvable systems from the general fifth order KdV type equation
\begin{equation}
u_t=u_{xxxxx}+a u^2u_x+b u_xu_{xx}+c u u_{xxx} \label{5KdV}
\end{equation}
with three arbitrary constants $a,\ b$ and $c$. It is known from the Painlev\'e analysis or the existence of higher order general symmetries, there exist three and only three integrable models from Eq. \eqref{5KdV}, namely, the Sawada-Kortera (SK), Kaup-Kupershmidt (KK) and fifth order KdV equations. In the following, we are interested in picking up these integrable systems again by finding CRE solvable systems from  \eqref{5KdV}.

Substituting \eqref{Ep} with \eqref{RCT} and $n=2$ (which is determined by the leading order analysis) into \eqref{5KdV}, we have
\begin{equation}
-2u_2w_xa_2[360a_2^4w_x^4+6u_2a_2^2w_x^2(b+2c) +au_2^2
]R^7+\sum_{i=0}^6K_iR^i=0,\label{K51}
\end{equation}
where $K_i\ (i=1,\ \ldots,\ 6)$ are complicated $w$-dependent but $R$-independent functions.

Vanishing the coefficient of $R^7$ in \eqref{K51} requires that $u_2$ must be proportional to $w_x^2$. Because the scaling of $u$ will not change the solvability of the model, we can simply take
\begin{equation}
u_2=-2a_2^2w_x^2,\label{kku2}
\end{equation}
without loss of generality and thus
\begin{equation}
a=3b+6c-90.\label{kka}
\end{equation}

Using the relations \eqref{kku2} and \eqref{kka}, \eqref{K51} is simplified to
\begin{equation}
-20w_x^5a_2^5(2b+3c-84)(2a_1a_2w_x^2+2a_2w_{xx}+u_1)
R^6+\sum_{i=0}^5K_iR^i=0.\label{K52}
\end{equation}
To eliminate the $R^6$ term in \eqref{K52},  two cases should be classified: (i) $2b+3c-84=0$, and (ii) $2b+3c-84\neq 0$. After tedious analysis, we find that six over-determined equations obtained from vanishing coefficients of $R^i$ for $i=0,\ 1,\ 2,\ \ldots,\ 5$ are not consistent for the first case. So we only need to consider the second case. In this latter case,  we have
\begin{equation}
u_1=-2a_2a_1w_x^2-2a_2w_{xx}.\label{kku1}
\end{equation}

Substituting \eqref{kku1} into \eqref{K52} yields
\begin{equation}
-8w_x^3a_2^5(b+c-30)[(a_1^2+8a_0a_2)w_x^4
+6w_x^2(a_1w_{xx}+u_0)+4w_xw_{xxx}-3w_{xx}^2]
R^5+\sum_{i=0}^4K_iR^i=0.\label{K53}
\end{equation}
To discussion further from \eqref{K53}, we find two nontrivial situations: (1) $b\neq 30 -c$, and (2) $b=30-c$.

{\bf Case (1): $b\neq 30 -c$.} In this case, we have
\begin{equation}
u_0=-\frac16(a_1^2+8a_2a_0)w_x^2-a_1w_{xx}
-\frac23S-\frac12\frac{w_{xx}^2}{w_x^2}.\label{kku0}
\end{equation}

Substituting \eqref{kku0} into \eqref{K53} leads to
\begin{equation}
4w_x^4a_2^4(2b+c-45)\left(\frac12w_x^2-S\right)_x
R^4+\sum_{i=0}^3K_iR^i=0.\label{K54}
\end{equation}

Because the coefficient of $R^4$ in \eqref{K54} is independence of $t$-derivative, this term can be canceled only for
\begin{equation}
c=45-2b.\label{kkc}
\end{equation}

After using the relation \eqref{kkc}, \eqref{K54} is simplified to
\begin{equation}
\frac13w_x^3a_2^3\left[(18-b)\big(40\delta w_{xx}^2-16S_{xx}+20\delta Sw_x^2-\delta^2w_x^4-4S^2\big)-12C\right]
R^3+\sum_{i=0}^2K_iR^i=0.\label{K55}
\end{equation}

Vanishing the coefficient of $R^3$ in \eqref{K55}, we have
\begin{equation}
C=\frac{18-b}{12}\big(40\delta w_{xx}^2-16S_{xx}+20\delta Sw_x^2-\delta^2w_x^4-4S^2\big).\label{kkC}
\end{equation}

With the help of \eqref{kkC}, we can find that the remaining terms of \eqref{K55} become
\begin{equation}
\left(75-4b\right)\left\{\frac{w_xa_2^2}3\left[
\delta^2w_x^4w_{xx}
-\delta\big(3S_{x}w_x^3+16Sw_{xx}w_x^2+15w_{xx}^3\big)
+w_x(S^2+2S_{xx})_x\right]
R^2+\sum_{i=0}^1K_iR^i\right\}=0.\label{K56}
\end{equation}

It is clear that except for
\begin{equation}
b=\frac{75}4,\label{kkb}
\end{equation}
the $w$ equation \eqref{kkC} is inconsistent with those of $K_2=K_1=K_0=0$. In conclusion, we obtain a CRE solvable system, which is just the known KK equation \cite{KK}
\begin{equation}
u_t=u_{xxxxx}+\frac{15}2uu_{xxx}
+\frac{75}4u_xu_{xx}+\frac{45}4u^2u_x. \label{KK}
\end{equation}
While the related CRE solvable theorem reads:

\bf Theorem 9. \rm If $w$ is a solution of
\begin{equation}
C=\frac{\delta^2}{16}w_x^2+\frac54\delta  w_x^2+S_{xx}+\frac52\delta w_{xx}^2+\frac14S^2,
\end{equation}
then
\begin{equation}
u=-\frac23\frac{w_{xxx}}{w_x}-a_1w_{xx}
-\frac16(8a_0a_2+a_1^2)w_x^2-2a_2(w_{xx}+a_1w_x^2)R
-2a_2^2w_x^2R^2
\end{equation}
is a CRE of the KK equation \eqref{KK}.

{\bf Case 2: $b=30-c$.} In this case, \eqref{K53} becomes
\begin{eqnarray}
&&12w_xa_2^4(15-c)\left\{w_x^3(S+2u_0)_x
+[(4a_0a_2+a_1^2)w_{xx}+2a_1S]w_x^4
+w_x^2w_{xx}(2S+3a_1w_{xx})+w_{xx}^3
\right\}
R^4\nonumber\\
&&\quad +\sum_{i=0}^3K_iR^i=0.\label{K54a}
\end{eqnarray}

The coefficient of $R^4$ in \eqref{K54a} shows that two further subcases should be considered: $c\neq 15$ and $c=15$.

{\bf Case 2.1: $c\neq 15$}. Vanishing the coefficient of $R^4$ of \eqref{K54a} results in
\begin{eqnarray}
u_0=\frac14\frac{w_{xx}^2}{w_x^2}-\frac12\frac{w_{xxx}}{w_x}
-a_1w_{xx}-\frac14(a_1^2+4a_0a_2)w_x^2+\lambda(t),\label{kdv5u0}
\end{eqnarray}
where $\lambda(t)\equiv \lambda$ is an integration function of $t$ and should be determined later by consistent conditions.

Due to the result \eqref{kdv5u0}, \eqref{K54a} becomes
\begin{eqnarray}
&&-\frac14w_x^3a_2^3\left\{(c-16)\delta^2w_x^4
-4\delta (3cS-2c\lambda-40S)w_x^2+
+4(c-12)(2S_{xx}-5\delta w_{xx}^2)\right.
\nonumber\\
&&\quad \left.+4S[S(c-16)+4c\lambda]+16(C-3c\lambda^2)
\right\}
R^3+\sum_{i=0}^2K_iR^i=0.\label{K55a}
\end{eqnarray}

Eliminating the $R^3$ term of \eqref{K55a} yields
\begin{eqnarray}
C=\frac{\delta^2}{16}(16-c)w_x^4
+\frac{\delta}4 (3cS-2c\lambda-40S)w_x^2
-\frac{c-12}4(2S_{xx}-5\delta w_{xx}^2)+3c\lambda^2+cS\lambda+4S^2-\frac{c}4S^2.\label{kdv5wt}
\end{eqnarray}

After substituting \eqref{kdv5wt} into
\eqref{K55a} and vanishing the remaining terms $R^i,\ i=2,\ 1,\ 0$, we find that the expansion \eqref{Ep} with $n=2$ is CRE only for
\begin{eqnarray}
c=10,\ \lambda_t=0.\label{kdv5c}
\end{eqnarray}
Otherwise two additional $w$ constraints have to be inserted. Therefore,  we have the  CRE theorem for the known fifth order KdV equation
\begin{equation}
u_t=u_{xxxxx}+30u^2u_x+20u_xu_{xx}+10uu_{xxx}.\label{KdV5}
\end{equation}

\bf Theorem 10. \rm If $w$ is a solution of
\begin{equation}\label{skdv5}
C=\frac38\delta^2w_x^4+\frac52\delta(S+2\lambda)w_x^2+S_{xx}
\frac52\delta w_{xx}^2+\frac32S^2+30\lambda^2+10\lambda S,
\end{equation}
then the CRE
\begin{equation}\label{ru_kdv5}
u=-\frac14(a_1^2+a_0a_2)w_x^2-a_1w_{xx}
-\frac12\frac{w_{xxx}}{w_x}
+\frac14\frac{w_{xx}^2}{w_x^2}+\lambda
-2a_2(a_1w_1^2+w_{xx})R-2a_2^2w_x^2R^2
\end{equation}
is a solution of the fifth order KdV equation \eqref{KdV5}.

{\bf Case 2.2: $c=15$.} Applying the similar analysis to this second subcase, we conclude that the only possible CRE model is just the so-called SK equation \cite{SK}
\begin{equation}
u_t=u_{xxxxx}+45u^2u_x+15(uu_{xx})_x.\label{SK}
\end{equation}
The corresponding CRE theorem is summarized as below.

\bf Theorem 11. \rm If $\{w,\ g\}$ is a solution of the consistent equations
\begin{subequations}\label{SKCRE}
\begin{equation}\label{SKwt}
C=S_{xx}+4S^2+\delta^2w_x^4-5\delta(S+3g) w_x^2-\frac52\delta w_{xx}^2+45g^2+30gS+15g_{xx},
\end{equation}
\begin{equation}\label{SKgxx}
\left(g_x^2+\delta g^2w_x^2-4g^3-2gg_{xx}-2g^2S\right)_x=0,
\end{equation}
\begin{eqnarray}\label{SKgt}
g_t&=&S_x(5g\delta w_x^2-9g^2-8gS)
+g_x\left[\delta^2w_x^4-\delta (5S-3g)w_x^2-9g^2-\frac52\delta w_{xx}^2-6gS+S_{xx}-3g_{xx}+4S^2\right]\nonumber\\
&&
-4\delta^2gw_{xx}w_x^3+3\delta g(3g+5S)w_xw_{xx}-gS_{xxx}+\frac{15}{2w_x}\delta gw_{xx}^3,
\end{eqnarray}
\end{subequations}
then the CRE
\begin{equation}
u=g-\frac13\frac{w_{xxx}}{w_x}-a_1w_{xx}
-\frac13(2a_0a_2+a_1^2)w_x^2-2a_2(w_{xx}+a_1w_x^2)R
-2a_2^2w_x^2R^2
\end{equation}
is a solution of the SK equation \eqref{SK}.

To guarantee the SK is CRE solvable, it is sufficient to take $g=0$ in \eqref{SKCRE}. However, the entrance of $g$ is still consistent because the consistent condition $g_{xxxt}=g_{txxx}$ of \eqref{SKgxx} and \eqref{SKgt} is nothing but \eqref{SKwt}.

All in all, from the above analysis, we have proved that there are only three CRE solvable systems of the form \eqref{5KdV} which are just three known integrable systems. Therefore, the conclusion is coincided with that from the Painlev\'e analysis and/or the existent conditions of high order symmetries.

\section{Common interaction behavior between soliton and cnoidal waves for CRE solvable systems}

Using the DT or BT related symmetry reduction approach, the interactions between a soliton and a cnoidal wave for the KdV equation, super-symmetric KdV, KP equations and AKNS systems have been discovered in \cite{Lou1,Lou2,Lou3,SolCN}. It is interesting that many different CRE solvable systems share similar interactions between a soliton and a cnoidal wave. Moreover, the $w$ solutions charactering the interactions between a soliton and a cnoidal wave for many different CRE solvable systems possess a common form
\begin{equation}\label{wsoliton}
w=k_1x+l_1y+\omega_1t+W(k_2x+l_2y+\omega_2t),
\end{equation}
where $W(k_2x+l_2y+\omega_2t)=W(\xi)\equiv W$ satisfies
\begin{equation}\label{Wsoliton}
W_{1\xi}^2=C_0+C_1W_1+C_2W_1^2+C_3W_1^3+C_4W_1^4,\ \qquad W_1\equiv W_\xi.
\end{equation}

It is clear that \eqref{Wsoliton} has explicit solutions expressed in terms of Jacobi elliptic functions.
If the interaction solution between a soliton and a cnoidal wave is allowed for a given CRE solvable model, the only difference lies in the relations among the constants
$C_0,\ C_1,\ C_2,\ C_3,\ C_4,\ k_1,\ k_2, \omega_1,\ \omega_2,\ l_1$ and $l_2$ ($l_2=l_1=0$ for 1+1 dimensional systems).

For simplicity, blow we just list the required relations among the constants for CRE solvable systems without details.

The KdV equation \eqref{KdV} asks for
\begin{eqnarray}
C_4&=&\delta,\nonumber\\
C_1&=&\frac{3k_2C_0}{k_1}
+\frac{2k_1\lambda +k_1^3\delta-2\omega_1}{k_2^3},\nonumber\\
C_2&=&\frac{3k_2^2C_0}{k_1^2}
+\frac{4\lambda +3k_1^2\delta}{k_2^2}
-\frac{\omega_2k_1+3k_2\omega_1}{k_1k_2^3},\nonumber\\
C_3&=&\frac{k_2^3C_0}{k_1^3}
+\frac{2\lambda}{k_2k_1}+\frac{3k_1\delta}{k_2}
-\frac{\omega_2k_1+\omega_1k_2}{k_2^2k_1^2},
\label{Ckdv}
\end{eqnarray}
while all the other constants remain free.

For the KP equation \eqref{KP}, there exist only three constraints among eleven constants,
\begin{eqnarray}
C_4&=&\delta,\nonumber\\
C_2&=&-\frac{3k_2^2C_0}{k_1^2}+\frac{2k_2C_1}{k_1}
+\frac{k_1^2\delta}{k_2^2}-\frac{(k_1^2l_2^2-k_2^2l_1^2)
\gamma}
{k_1^2k_2^4}
-\frac{\omega_2k_1-k_2\omega_1}{k_1k_2^3},\nonumber\\
C_3&=&-\frac{2k_2^3C_0}{k_1^3}+\frac{k_2^2C_1}{k_1^2}
+\frac{2\delta k_1}{k_2}
-\frac{\omega_2k_1-\omega_1k_2}{k_2^2k_1^2}-\frac{2(k_1^2l_2^2-k_2^2l_1^2)\gamma}
{k_1^3k_2^3}.
\label{Ckp}
\end{eqnarray}

The Boussinesq system \eqref{Bq} has the constraints as
\begin{eqnarray}
C_4&=&\delta,\nonumber\\
C_2&=&-\frac{3k_2^2C_0}{k_1^2}+\frac{2k_2C_1}{k_1}
+\frac{k_1^2\delta}{k_2^2}
-\frac{\omega_2^2k_1^2-k_2^2\omega_1^2}{k_1^2k_2^4},\nonumber\\
C_3&=&-\frac{2k_2^3C_0}{k_1^3}+\frac{k_2^2C_1}{k_1^2}
+\frac{2\delta k_1}{k_2}
-\frac{2(2k_1^2\omega_2^2-k_2^2\omega_1^2
-k_1k_2\omega_1\omega_2)}
{3k_1^3k_2^3}.
\label{Cbq}
\end{eqnarray}

For the SK model \eqref{SK}, the constant constraints can be written as
\begin{eqnarray}
C_4&=&-\delta,\nonumber\\
C_1&=&\frac{k_1}{k_2^3}(2C_2k_2^2-3C_3k_1k_2
-4\delta k_1^2),\nonumber\\
\omega_1&=&-9\delta^2k_1^5-\frac32k_1k_2
\delta(5C_3k_1^3-2C_2k_2k_1^2+10C_0k_2^3)
-\frac{k_2^3}2(9C_0C_3k_2^2-2k_1C_2^2k_2+3C_2C_3k_1^2),\nonumber\\
\omega_2&=&45\delta^2k_2k_1^4+3k_2^2(C_0k_2^3-5C_2k_2k_1^2+14C_3k_1^3)\delta
+k_2^3(C_2k_2-3k_1C_3)^2.
\label{Csk}
\end{eqnarray}

For the KK system \eqref{KK}, we have the following constant constraints
\begin{eqnarray}
C_4&=&\delta,\nonumber\\
C_1&=&\frac{4k_1^3\delta}{k_2^3}+\frac{k_1}{k_2^2}
(2C_2k_2-3C_3k_1),\nonumber\\
C_0&=&\frac{3k_1^4\delta}{k_2^4}+\frac{k_1^2}{k_2^3}
(C_2k_2-2C_3k_1),\nonumber\\
\omega_1&=&\frac94\delta^2k_1^5+\frac34k_1^3k_2
\delta(C_2k_2-3C_3k_1)+\frac{k_1k_2^2}{16}(C_2k_2-3C_3k_1)^2,
\nonumber\\
\omega_2&=&\frac94\delta^2k_2k_1^4+
\frac34k_1^2k_2^2\delta(C_2k_2-3C_3k_1)
+\frac{k_2^3}{16}(C_2k_2-3k_1C_3)^2.
\label{Ckk}
\end{eqnarray}

For the usual fifth order KdV equation \eqref{KdV5}, there are two sets of constant constraints. The first set reads
\begin{eqnarray}
C_4&=&\delta,\nonumber\\
C_0&=&\frac{k_1}{k_2^4}(C_1k_2^3-C_2k_1k_2^2
+C_3k_1^2k_2-\delta k_1^3) ,\nonumber\\
\omega_1&=&30\lambda^2k_1+\lambda(15C_2k_1k_2^2+10\delta k_1^3-10 C_1 k_2^3-15C_3k_2k_1^2)-2C_1C_3k_1k_2^4+5C_1\delta k_1^2k_2^3\nonumber\\
&&+\frac{13}4C_2C_3k_1^2k_2^3
-\frac{15}2C_2\delta k_2^2k_1^3
+\frac{19}2\delta C_3k_1^4k_2
+\frac{1}2C_2C_1k_2^5
-\frac{13}2\delta^2k_1^5
-\frac{21}8C_3^2k_1^3k_2^2
-\frac{5}8C_2^2k_1k_2^4,
\nonumber\\
\omega_2&=&-\frac{k_2}8[k_2^4(4C_1C_3-3C_2^2)-2k_1k_2^3(8\delta C_1-5C_2 C_3)-k_1^2k_2^2(15C_3^2+4C_2\delta)+44\delta k_1^3(k_2C_3-\delta k_1)]\nonumber\\
&&+30k_2\lambda^2-5k_2\lambda(C_2k_2^2-3k_1k_2C_3
+6\delta k_1^2),
\label{Ckdv5}
\end{eqnarray}
and the second is given by
\begin{eqnarray}
C_4&=&\frac{\delta}9,\qquad C_3=\frac{4\delta k_1}{9k_2},\quad C_2=\frac{2(\delta k_1^2-6\lambda)}{3k_2^2},\nonumber\\
C_0&=&\frac{k_1}{3k_2^4}(12k_1\lambda+3C_1k_2^3-\delta k_1^3),\nonumber\\
\omega_1&=&\frac{16}3\lambda \delta k_1^3-12C_1\lambda k_2^3 -40k_1\lambda^2,\quad \omega_2=56\lambda^2k_2.
\label{Ckdv52}
\end{eqnarray}

For the AKNS system \eqref{AKNS}, nine arbitrary constants $C_i,\ i=0,1,\ldots, 4,\ k_1,\ k_2,\ \omega_1,\ \omega_2$ should satisfy three constraints
\begin{eqnarray}
C_4&=&\delta,\nonumber\\
C_3&=&\frac{2k_1\delta}{k_2}+\frac{C_1k_2^2}{k_1^2}
-\frac{2C_0k_2^3}{k_1^3}
+\frac{8(b\lambda k_1+\omega_1) (k_1\omega_2-k_2\omega_1)}{k_2^2 k_1^3},\nonumber\\
C_2&=&\frac{k_1^2\delta}{k_2^2}+\frac{2C_1k_2}{k_1}
-\frac{3C_0k_2^2}{k_1^2}
+\frac{4(2b\lambda k_1k_2+3k_2\omega_1-k_1\omega_2) (k_1\omega_2-k_2\omega_1)}{k_2^4 k_1^2}.
\label{CAKNS}
\end{eqnarray}

For the sG equation \eqref{sgu}, there are five constraints in the form of
\begin{eqnarray}
C_0&=&\frac{k_1^2\omega_1^2\delta}
{k_2^2\omega_2^2}-\frac{2mk_1\omega_1
(k_1k_2\lambda^2-\omega_1\omega_2)}
{\omega_2^2k_2^2\lambda
(k_1\omega_2-k_2\omega_1)},\nonumber\\
C_1&=&\frac{2\delta k_1\omega_1}{k_2^2\omega_2^2}(k_1\omega_2+k_2\omega_1)
-\frac{2m\lambda k_1(k_1\omega_2+2k_2\omega_1)}
{k_2\omega_2^2(k_1\omega_2-k_2\omega_1)}
+\frac{2m\omega_1(2k_1\omega_2+k_2\omega_1)}
{k_2^2\omega_2\lambda(k_1\omega_2-k_2\omega_1)},\nonumber\\
C_2&=&\frac{\delta} {k_2^2\omega_2^2}(k_1^2\omega_2^2+k_2^2\omega_1^2
+4k_1k_2\omega_1\omega_2)
-\frac{2m\lambda(k_2\omega_1+2k_1\omega_2)}
{\omega_2^2(k_1\omega_2-k_2\omega_1)}
+\frac{2m(2k_2\omega_1+k_1\omega_2)}
{k_2^2\lambda(k_1\omega_2-k_2\omega_1)},\nonumber\\
C_3&=&\frac{2\delta}{k_2\omega_2}(k_1\omega_2+k_2\omega_1)
-\frac{2m(k_2^2\lambda^2-\omega_2^2)}{k_2\omega_2\lambda
(k_1\omega_2-k_2\omega_1)},\nonumber\\
C_4&=&\delta,
\label{CsG}
\end{eqnarray}
while all the other constants $\omega_1,\ \omega_2, \ k_1,\ k_2,\ \lambda,$ and $ \delta$ are free.

The VN system \eqref{VN} requires four constraints
\begin{eqnarray}
C_4&=&\delta,\nonumber\\
C_3&=&\frac{k_2l_2C_1}{k_1l_1}-\frac{(k_2l_1+k_1l_2)(C_0l_2^2k_2^2-\delta k_1^2l_1^2)}{k_1^2l_1^2k_2l_2},\nonumber\\
C_2&=&\frac{k_1l_1\delta}{k_2l_2}
+\left(\frac{k_2}{k_1}+\frac{l_2}{l_1}\right)C_1-
\left(\frac{k_2^2}{k_1^2}+\frac{k_2l_2}{k_1l_1}
+\frac{l_2^2}{l_1^2}\right)C_0,\nonumber\\
\omega_2&=&\frac{k_2\omega_1}{k_1}-\frac{k_2(k_1l_2-k_2l_1)
[C_0l_2k_2^2(k_1l_2+2k_2l_1)
-l_1k_1(C_1l_2k_2^2-\delta l_1k_1^2)]}{4l_2l_1^2k_1^2}.
\label{Cvn}
\end{eqnarray}

It should be pointed out that not all the CRE solvable systems possess interaction solutions between a soliton and a cnoidal wave. For instance, two C-integrable systems,  the Burgers equation \eqref{Be} and the DWW system \eqref{DWW}, do not have the $w$ solution in the form of \eqref{wsoliton}. In fact, it is well known that for the Burgers equation, even the single cnoidal wave solution is not allowed.

\section{Summary and discussions}

In summary, with the help of the Riccati equation, we have proposed a simple CRE method by which many kinds of nonlinear systems can be solved. It is found that various integrable models such as the KdV, fifth KdV, KP, Boussinesq, AKNS, NLS, sG, SK, KK, DWW, VN and Burgers models are CRE solvable. Though not all the integrable systems are CRE solvable, it is strongly indicated that the CRE solvable systems are integrable. Therefore, it is feasible to use the CRE method to detect CRE solvable systems first and then one can go further to check other integrable properties. This idea is applied to a general fifth order KdV type system, and it is demonstrated that the only possible CRE solvable systems are the SK, KK and the usual fifth order KdV equations, which coincides with the integrability classifications via the Painlev\'e analysis and the existence of higher order symmetries.

In fact, only to pick out CRE solvable models from general systems, one may use some special solutions of the Riccati equation. For instance, taking $a_1=0, \ a_0=1$ and $a_2=-1$, the Riccati equation \eqref{RCT} possesses the solution $R=\tanh (w)$, and accordingly, the simplified CRE can be termed as consistent tanh expansion (CTE). Obviously, the CRE solvable systems are CTE solvable, and vice versa.

For many CRE solvable systems, there is a common interesting exact interaction solution between a soliton and a cnoidal periodic wave, determined by Eqs. \eqref{wsoliton} and \eqref{Wsoliton}. The only differences are the constant constraints related to the possible dispersion relations and average backgrounds. The detailed interactions between solitons and cnoidal waves for the KdV equation, the KP equation and the AKNS system have been discussed in several references \cite{Lou1,Lou2,Lou3,SolCN}. The more about the method and the associated $w$ equations needs further study.

\section*{Acknowledgement.}
The author is in debt to the helpful discussions with Drs. Y. Q. Li, X. Y. Tang, C. L. Chen, Profs. X. B. Hu, R. X. Yao, Q. P. Liu and Y. Chen.
The work is sponsored by the National Natural Science Foundations of China (Nos. 11175092, 11275123, 11205092 and 10905038), Shanghai Knowledge Service Platform for Trustworthy Internet of Things (No. ZF1213), Talent Fund and K. C. Wong Magna Fund in Ningbo University.

\small{
}

\begin{thebibliography}{99}
\bibitem{IST}Gelfand I M and Levitan B M 1955 Amer. Math. Soc. Trans. \bf 1 \rm 253\\
Ablowitz M and Segur H 1981 \em Solitons and the Inverse Scattering Transform \rm SIAM, Philadelphia\\
Gardner C S, Greene J M, Kruskal M D and Miura R M 1967 Phys. Rev. Lett. \bf 19 \rm 1095
\bibitem{DT}Gu C H, Hu H S and Zhou Z X 2005 \em Darboux Transformations in Integrable Systems
Theory and their Applications to Geometry, \rm
Series: Mathematical Physics Studies, Vol. 26 Springer, Dordrecht
\bibitem{NL}Cao C W 1987 Henan Sci. \bf 5 \rm 1\\
Cao C W 1988 Chin. Q. J. Math. \bf 3 \rm 90\\
Cao C W 1990 Sci. China A \bf 33 \rm 528\\
Cao C W and Geng X G 1990 J. Phys. A \bf 23 \rm 4117\\
Cao C W and Geng X G 1991 J. Math. Phys. \bf 32 \rm 2323
\bibitem{tanh}Lan H B and Wang K L 1990 J. Phys. A: Math. Gen. \bf 23 \rm 3923\\
    Lou S Y, Huang G X and Ruan H Y 1991 J. Phys. A: Math. Gen. \bf 24 \rm L587\\
    Parkes E J and Duffy B R 1996 Comp. Phys. Commun. \bf 98 \rm 288\\
    Fan E G 2000 Phys. Lett. A \bf 277 \rm 212
\bibitem{Ric}Wang Q, Chen Y and Zhang H Q 2005 Chaos Solitons and Fractals \bf 25 \rm 1019\\
    Li B, Chen Y, Xuan H N and Zhang H Q 2004 Appl. Math. Comp. \bf 152 \rm 581
\bibitem{homo}Wang M L, Zhou Y B and Li Z B 1996 Phys. Lett. A \bf 216 \rm 67
\bibitem{Lou1}Cheng X P, Chen C L and Lou  S Y
\em Interactions among different types of nonlinear waves described by the Kadomtsev-Petviashvili Equation \rm
arXiv:1208.3259
\bibitem{Lou2}Lou S Y, Cheng X P, Chen C L and Tang X Y \em Interactions Between Solitons and Other Nonlinear Schr\"odinger Waves     \rm arXiv:1208.5314
\bibitem{Lou3}Gao X N, Lou S Y and Tang X Y 2013 JHEP \bf 1305 \rm 029 DOI: 10.1007/JHEP05(2013)029
\bibitem{KdV}Korteweg D J and de Vries F 1895
Philos. Mag. \bf 39 \rm 422 \\
Crighton D G 1995 Acta Appl. Math. \bf 39 \rm 39
\bibitem{KP}Kadomtsev B B and Petviashvili V I 1970 Dokl.
Akad. Nauk SSSR \bf 192 \rm 753
\bibitem{Bq}Boussinesq J de 1871 Comptes Rendus Acad. Sci. Paris \bf 72 \rm 755 \\
    Zakharov V E 1973 JETP \bf 65 \rm 219
\bibitem{AKNS}Kivshar Y S and Malomed B A 1989 Rev. Mod. Phys. \bf 61 \rm 763\\
Kivshar Yu S and Luther-Davies B 1998 Phys. Rep. \bf 298 \rm 81
\bibitem{sG} Ablowitz M J, Kaup D J, Newell A C and Segur H 1973 Phys. Rev. Lett. \bf 30 \rm 1262
\bibitem{VN} Bogdanov L V 1987 Theor. Math. Phys. \bf 70 \rm 219
\bibitem{DWW} Kupershmidt B A 1986 Mech. Res. Commun. \bf 13 \rm 47
\bibitem{BE}Burgers J M 1948  Adv. Appl. Mech. \bf 1 \rm 171\\
Cole J D 1950 Q. Appl. Math. \bf 9 \rm 225\\
Hopf E 1950 Comm. Pure and Appl. Math. \bf 3 \rm 201
\bibitem{KK}Kaup D J 1980 Stud. Appl. Math. \bf 62 \rm 189
\bibitem{SK}Sawada K and Kotera T 1974 Progr. Theor. Phys. \bf 51 \rm 1355 \\
Caudrey P J, Dodd R K and Gibbon J D 1976  Proc. Roy. Soc. London A \bf 351 \rm 407
\bibitem{SolCN}Lou S Y and Hu X B 1993 Chin. Phys. Lett. \bf 10 \rm 577 \\
    Lou S Y, Hu X R and Chen Y 2012 J. Phys. A: Math. Theor. \bf 45 \rm 155209\\
    Hu X R, Lou S Y and Chen Y 2012 Phys. Rev. E \bf 85 \rm 056607
\end{thebibliography}
\end{document}